# Fabrication of Rare Earth-Doped Transparent Glass Ceramic Optical Fibers by Modified Chemical Vapor Deposition

Wilfried Blanc[1,*], Valérie Mauroy[1], Luan Nguyen[2], Shivakiran Bhaktha B.N.[1], Patrick Sebbah[1], Bishnu P. Pal[3], and Bernard Dussardier[1]

[1] *LPMC, Université de Nice Sophia Antipolis, CNRS UMR6622, Parc Valrose, 06108 Nice Cedex 2, France*

[2] *CRHEA, CNRS UPR10, rue B. Gregory, Les Lucioles 1, 06560 Valbonne, France*

[3] *Physics Department, Indian Institute of Technology Delhi, New Delhi 110 016, India*

[*]*Corresponding author: wilfried.blanc@unice.fr*

Rare earth (RE) doped silica-based optical fibers with transparent glass ceramic (TGC) core was fabricated through the well-known modified chemical vapor deposition (MCVD) process without going through the commonly used stage of post-ceramming. The main characteristics of the RE-doped oxyde nanoparticles namely, their density and mean diameter in the fibers are dictated by the concentration of alkaline earth element used as phase separating agent. Magnesium and erbium co-doped fibers were fabricated. Optical transmission in term of loss due to scattering as well as some spectroscopic characteristics of the erbium ions was studied. For low Mg content, nano-scale particles could be grown with and relatively low scattering losses were obtained, whereas large Mg-content causes the growth of larger particles resulting in much higher loss. However in the latter case, certain interesting alteration of the spectroscopic properties of the erbium ions were observed. These initial studies should be useful in incorporating new doped materials in order to realize active optical fibers for constructing lasers and amplifiers.

# I. Introduction

Developing of new rare-earth (RE)-doped optical fibers for power amplifiers and lasers requires continuous improvements in the fiber absorption and amplification properties. Silica glass as a fiber host material has proved to be very attractive. However some potential applications of RE-doped fibers suffer from limitations in terms of spectroscopic properties that may result from clustering and/or inappropriate local environment of RE ions in silica host. To overcome those limitations, the route of interest here consists of embedding the amplifying RE ions within oxide nanoparticles (NP) of composition and structure different from those of silica: this would provide a beneficial local environment to RE ions in terms of their spectroscopic properties.

In the literature, the term Transparent Glass Ceramics (TGC) is used to refer to crystalline NP embedded into amorphous glass [1]. However, the NP may also be amorphous, such as those obtained through phase separation [2]. In the following we use the term TGC for both cases for convenience.

One possible drawback of TGC in optical fibers may be due to scattering loss caused by NP. The acceptable size range of the NP is strongly application-dependent. For conventional fiber lasers and amplifiers with high gain per unit meter, they should be small enough so as to keep scattering loss within acceptable limits, typically < 1dB/m [3]. On the contrary, random fiber lasers require a high level of scattering, for which bigger NP are desirable [4].

The very few reports on RE-doped TGC single-mode fibers are based on low melting mixed oxides prepared by a rod-in-tube technique [5], or mixed oxyfluorides using a double-crucible technique [6]. They both require implementing a subsequent ceramming stage close to the glass transition temperature. This heat treatment stage may greatly weaken the fibers, and it introduces certain complexicities in the global fabrication process. Further, the compatibility of these materials with silica-based fiber components is a serious issue due to their low melting point and low damage threshold (especially in high optical power applications). Only one report deals with silica-based TGC fiber preforms prepared by the well-known MCVD (modified chemical vapor deposition) technique: transition metal doped NP was synthesized before being incorporated in the preform through use of a slurry method [7].

We have proposed a more straightforward technique which allowed embedding of RE ions within the *in-situ* grown oxide NP in silica-based preforms prepared by MCVD, and without the need of a ceramming stage [8]. The technique exploited the spontaneous phase separation process in silicate systems when they contain alkaline-earth elements [2]. Two key advantages of this process were that (i) NPs are grown *in-situ* during the course of the fabrication process and (ii) there is no need (and associated potential risks) for manipulation of NPs by an operator. Further, the process takes advantage of the high compositional control and purity typical of the MCVD technique. The role of CaO in the formation of NP in fibers has been already reported [8, 9]. Their typical diameter was more than 100 nm, causing high scattering losses. In this letter, we report on the fabrication for the first time of RE-doped fibers by the MCVD process with a TGC core. We focus on the incorporation of Mg, which was demonstrated very recently to lead to phase separation in silicate glass based fibers, even under a rapid cooling yield [10]. Results reported demonstrate the potentiality of this fabrication process for applications such as fiber amplifiers and lasers.

## II. Experimental Procedure

Preforms were fabricated by the conventional MVCD technique [11]. The so-called 'solution doping technique' [12] was applied to incorporate magnesium and erbium ions: the core porous layer is soaked with an alcoholic solution of $ErCl_3:6H_2O$ and $MgCl_2:6H_2O$ of desired concentrations. After drying of the solvent, the core layer is sintered down to a dense glass layer. Then the tube is collapsed into a solid rod, referred to as perform, at an elevated temperature higher than 1800 °C. Preforms were stretched into 125-µm fibers using a draw tower at temperatures higher than 2000 °C under otherwise normal conditions. The fiber core diameters are about 8 µm. To raise the core refractive index and ease the fabrication, germanium (1.85 mol%) and small amounts of phosphorus (0.8 mol%) were added. A set of fabricated preforms showed a refractive index contrast $\Delta n$ in the range of 1.7 to $4.10^{-3}$. The critical process parameter is the concentration of magnesium in the soaking solution. Here we report on samples labeled as Fiber A and Fiber B, doped with solutions containing 0.1 and 1 mol/l of Mg salts, respectively. The resulting Mg concentration in Fiber A was 0.1 mol%. The erbium concentration of 0.01 mol/l was kept constant in the solution. The attenuation coefficient due to absorption by $Er^{3+}$ ions at 1.53 µm was measured to be 9 dB/m in Fiber A. From this value and the estimated overlap between the guided mode and the doped region, the erbium ions concentration was estimated to ~100 ppm.

## III. Results and discussion

*(1) DNP characterizations*

The fibers were characterized through scanning electron microscopy (SEM) in the backscattered electrons mode. Typical SEM pictures from the exposed core section of cleaved fibers (Fibers A and B) are shown on Figs. 1a and 1b. The gray disk corresponds to the fiber core. The dark central part of the core is caused by the evaporation of germanium element; this is a common artifact of the MCVD technique that can be corrected through process optimization. Nanoparticles are only observed when Mg is added, they are visible as bright spots. They show an important compositional contrast compared to the silica background. The statistical histogram of the size distribution of the NP for Fiber A is presented in Fig. 1a. The mean particle size is 48 nm and no particle bigger than 100 nm was observed in these MgO-doped fibers, unlike the case of CaO-doped fibers reported elsewhere [13]. Analysis of the SEM images also revealed that the inter-particle distance is in the 100 ~ 500 nm range. When the solution concentration increases from 0.1 to 1 mol/l, the inter-particle distance remains nearly the same but the mean particle diameter almost doubles to reach 76 nm (Fig. 1a & 1b). In Fiber B, NPs up to 160 nm were observed (Fig. 1b). Magnesium is known to be a network modifier in silica glass [10, 14]. By increasing the Mg concentration, the glass structure becomes more 'open' for the cations to move through it and hence would facilitate formation of bigger particles.

Although the size distributions observed in both the fibers are not identical, the total volume of the particles varies accordingly to the Mg concentration in the doping solution. Indeed, the total volume of NPs in Fiber A and B were $7\times10^6$ and $73\times10^6$ nm$^3$ (as seen from Figs. 1a and 1b), respectively. It is proposed that all Mg, and part or all Er ions, in the core are hosted within the NPs as observed in Ca-doped silica preforms by us [13]. As a consequence, one expects that a high content of Mg is in the NPs, and part or all Er ions would experience a local environment different from that of silica.

*(2) Spectroscopic characterizations of $Er^{3+}$*

The emission spectra and lifetime from Fibers A and B were measured at room temperature around 1.55 µm under 980-nm pump excitation. The emission spectrum from Fiber A (Fig. 2) is similar to that in a silicate environment. Fiber B shows a distinct broadening of its spectrum (FWHM is 44 nm) by as much as ~ 50 % compared to Fiber A. In comparison, more than 10 at% of aluminum, as network modifier, would have been necessary to obtain the same FWHM in 'type III' Er-doped fibers in optical amplifiers for telecommunications [15], because Al and Er are evenly distributed across the core volume. Moreover, the shape of the fluorescence spectrum from Fiber B is quite unusual: it decreases monotonically between the peaks at 1.53 and 1.55 µm and the commonly observed dip at 1.54 µm is absent. These features would be attractive for realizing intrinsically gain flattened fiber amplifiers, provided sufficient minimization of the scattering loss is ensured through process optimization. Fluorescence decay curves are reported on Fig. 2. Both the fibers A and B produced decaying fluorescence curves which were fitted with a single-exponential. The measured lifetimes were 11.7 and 6.7 ms, respectively.

Modifications of the $Er^{3+}$ spectroscopic properties in the TGC optical fibers are clearly evident from the above results when the Mg concentration increases. Such observations were previously reported in Ca-doped preforms where a high content of Ca and P were measured in NPs [13]. Here this is observed in optical fibers for the first time. The broadening of the fluorescence curve and the lower lifetime obtained from Fiber B is tentatively interpreted as an effect of the modification of the erbium ions averaged local field. In other words, erbium ions in Fiber B are, on average, located in a medium with higher local field compared to Fiber A. This induces stronger spontaneous transition probability, and hence a shorter lifetime [16]. Although the exact composition of the NPs is not yet known, emission spectrum and fluorescence lifetime of erbium ions in Fiber B are closely related to the results reported in the literature in phosphate glasses [17, 18]. When Mg concentration increases, erbium ions environment changes from silicate to phosphate. Further composition analyses are under way with Nanosims-50 analyses to determine the partition of $Er^{3+}$ ions and the composition of the NPs. Moreover, we intend to make further spectroscopic studies, such as Resonant Fluorescence Line Narrowing, for more detailed report elsewhere at later date.

*(3) Attenuation measurements*

The scattering loss in fibers A and B were measured through the standard cut-back method. The fiber-bend radius was kept > 20 cm to minimize bend loss. The attenuation spectrum of fiber A is displayed in Fig. 3. At the wavelength of 1350 nm, losses were measured to be 0.4 dB/m only. This value is comparable with the attenuation measured in low-melting temperature transparent glass ceramic fibers [5] and is compatible with amplifier applications. Increase in loss with decrease in wavelength is attributable to light scattering induced by the NPs. Due to the small particle size we assume Rayleigh scattering is the dominant source of scatter loss, which was estimated according to the formula: $\alpha$ (dB/m) = $4.34 \times C_{Rayleigh}.N.\Gamma$, where $N$ is the DNP density (m$^{-3}$), $\Gamma$ is the overlap factor between the field and the core containing the DNP ($\Gamma$ = 0.3 at 1350 nm in Fiber A) and $C_{Rayleigh}$ (m$^2$) is the Rayleigh scattering coefficient defined as follows [19]:

$$C_{Rayleigh} = \frac{(2\pi)^5}{48} \times \frac{d^6}{\lambda^4} \times n_m^4 \times \left( \frac{n_n^2 - n_m^2}{n_n^2 + 2n_m^2} \right)^2 \qquad (1)$$

where $d$ is the nanoparticle diameter, $n_m$ and $n_n$ represent refractive indices of the host material and particles, respectively. The actual particle composition is not known and two refractive indices were considered according to the previous discussion about erbium ions environment: 1.65 and 1.53, like that of Mg-based oxide such as Mg$_2$SiO$_4$ [20], and phosphate glasses [18], respectively. Assuming $d$ = 48 nm and $n_m$ = 1.45, fitting of the experimental data with Eq. 1 yields a particle density $N$ (the only adjustable parameter) of $1.2 \times 10^{19}$ and $6.8 \times 10^{19}$ particles/m$^3$ for Mg$_2$SiO$_4$ and phosphate glasses refractive indices, respectively. The coefficient of determination, R$^2$, is 0.98134. Due to the adjustment of the NP density as free parameter, both fitting curves are superposed, and only the Calculated Rayleigh scattering curve for $n_n = 1.65$ is shown. The fitted mean inter-particle distance is 437 and 244 nm for both refractive indices, respectively, in agreement with the 100 ~ 500 nm-range estimated from the SEM pictures from cleaved fibers.

The difference between the experimental data and Rayleigh scattering curve beyond 1350 nm is attributable to bending loss alone. Indeed the normalized frequency $V$ at 1350 nm is < 1.3 ($LP_{11}$ mode cut-off wavelength is 700 nm). It is worth mentioning that, for practical application, our fabrication technique allows any necessary waveguide optimization without preventing the TGC growth. The attenuation of Fiber B was extremely high, more than 100 dB/m. It is attributed to the presence of the bigger sized particles (see Fig. 1b). According to Eq.1 and the Rayleigh formula, 120-nm particles would induce a loss 100 dB/m for $n_n = 1.65$. This shows that the particle mean size must be less than 50 nm for potential applications of this kind of fibers as fiber lasers and amplifiers..

**IV. Summary**

A method to fabricate $Er^{3+}$-doped TGC fibers entirely through the MCVD process is demonstrated. By adding magnesium to the silica-based composition, NPs of 40 nm in diameter are obtained through *in situ* growth without requiring a separate process to realize NP (such as post-process ceramming). A low-loss fiber is reported for the first time with this technique. An important result of this study is that the type of obtained nanoparticle by this technique could be as large as ~ 50 nm for applications such as in fiber amplifiers and lasers. Moreover, a broadening of the emission spectrum by as much as ~ 50% is observed with attractive features to realize gain flattened fiber amplifiers. This important feature should be useful through more careful process optimization to control the particle size. More generally, this concept might have great potentials as possible solutions to address various current issues in amplifying fibers, including (but not limited to) realization of intrinsically gain flattened amplifiers, etc.

**Acknowledgement**

This work was partially supported by CNRS (France), the Ministère des Affaires Etrangères (France) and Department of Science and Technology (India) through an Indo-French Research Network 'P2R' program. LPMC is a member of the GIS 'GRIFON' (http://www.unice.fr/GIS/). The authors thank Michèle Ude and Stanislaw Trzésien for preparation of the samples.

**Figure Captions**

Fig. 1: SEM pictures of a Mg-doped fiber and histograms of the particle size (a: Fiber A, b: Fiber B).

Fig. 2: Room temperature fluorescence emission spectra from Fibers A (full line) and B (dotted line). The Mg concentrations in the doping solution were 0.1 and 1.0 mol/l, respectively. Excitation wavelength was 980 nm. Spectra are normalized at their maximal intensity. Insert: Fluorescence decay from Fibers A (square: experimental data, full line: single exponential fitting) and B (triangle: experimental data, dotted line: single exponential fitting) measured at room temperature and at 1.5 µm. Excitation wavelength was 980 nm.

Fig. 3: Transmission spectrum of a Mg-doped fiber (Fiber A). Dots: experimental data, full line: as calculated based on Rayleigh scattering.

Fig 1

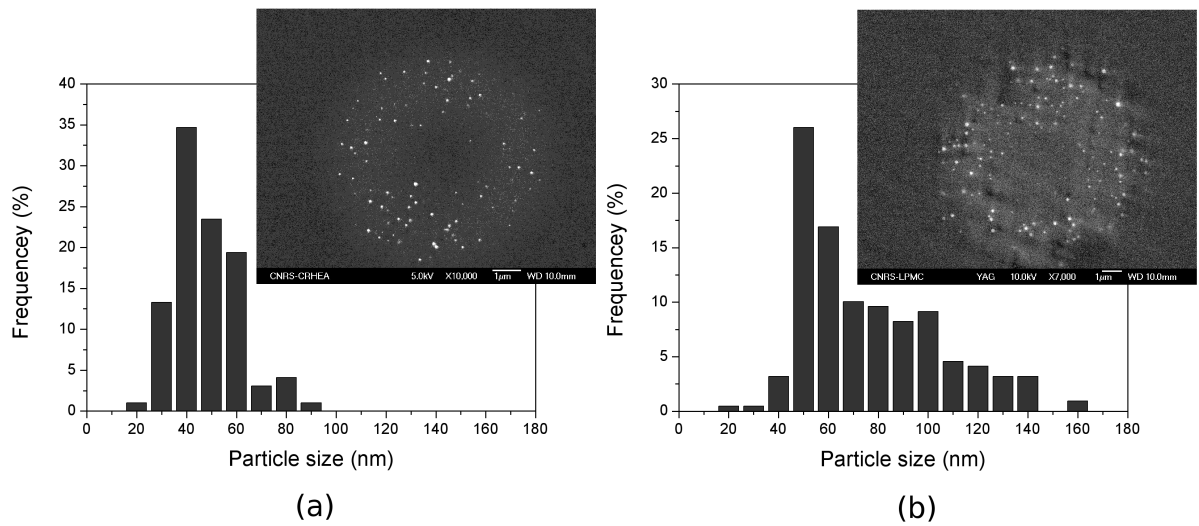

(a)     (b)

Fig 2

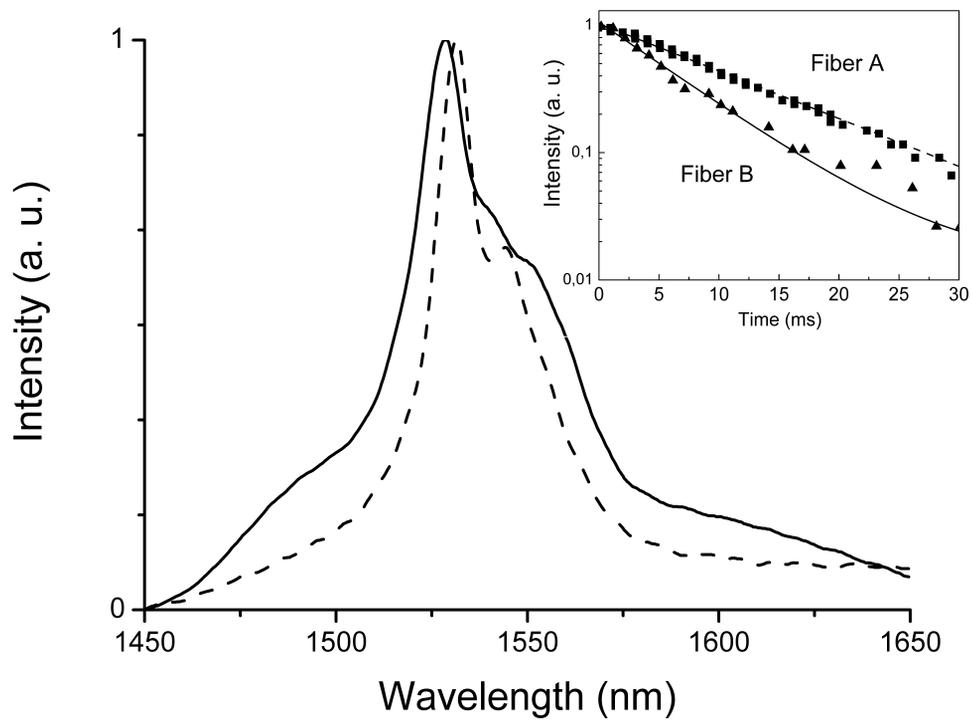

Fig. 3

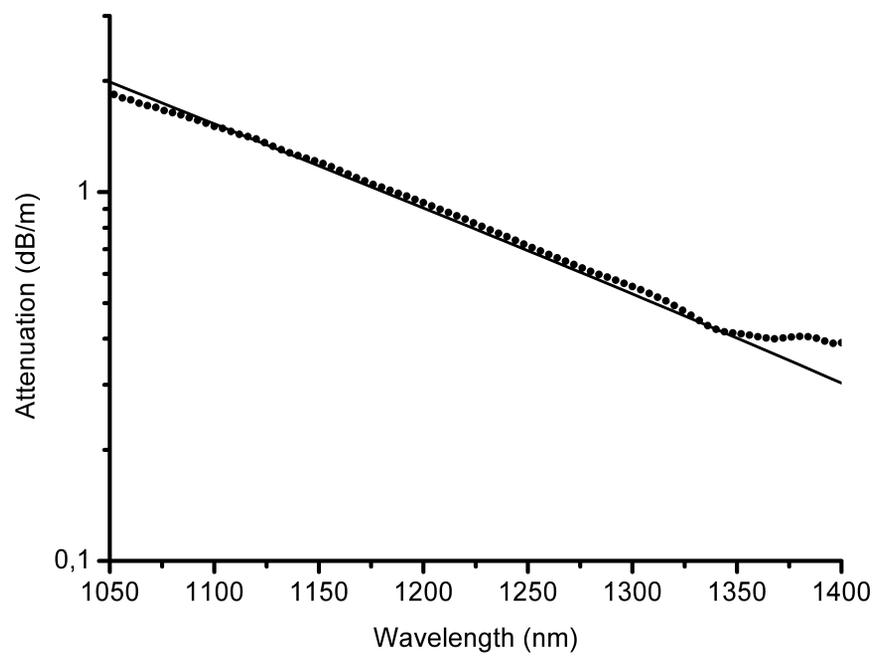